\documentclass[%
 reprint,
 amsmath,amssymb,
 aps,
prb,
]{revtex4-1}

\usepackage{amssymb}
\usepackage{fullpage}
\usepackage{graphicx}
\usepackage{bibunits}
\usepackage{wrapfig}
\usepackage{floatflt}
\usepackage{color}
\usepackage{xspace}
\usepackage{dsfont}
\usepackage[official,right]{eurosym}
\usepackage[top=0.8in, bottom=0.8in, left=0.75in, right=0.75in]{geometry}
\usepackage[utf8]{inputenc}

\newcommand{\PbV}{Pb$_2$VO(PO$_4$)$_2$\xspace}

\newcommand{\dd}{\mathrm{d}}

\newcommand{\be}{\begin{equation} }
\newcommand{\ee}{\end{equation} }
\newcommand{\bea}{\begin{eqnarray} }
\newcommand{\eea}{\end{eqnarray} }

\usepackage{graphicx}% Include figure files
\usepackage{dcolumn}% Align table columns on decimal point
\usepackage{bm}% bold math
\usepackage{hyperref}% add hypertext capabilities
\makeatletter
\AtBeginDocument{\let\LS@rot\@undefined}
\makeatother

\begin{document}

\title{Pre-saturation phase in the frustrated ferro-antiferromagnet \PbV}

\author{F. Landolt}
\email{landoltf@phys.ethz.ch}
\affiliation{Laboratory for Solid State Physics, ETH Z\"urich, 8093 Z\"urich, Switzerland}

\author{S. Bettler}
\altaffiliation[Present address: ]{Department of Physics and Materials Research Laboratory, University of Illinois, Urbana, IL 61801, USA}
\affiliation{Laboratory for Solid State Physics, ETH Z\"urich, 8093 Z\"urich, Switzerland}

\author{Z. Yan}
\affiliation{Laboratory for Solid State Physics, ETH Z\"urich, 8093 Z\"urich, Switzerland}

\author{S. Gvasaliya}
\affiliation{Laboratory for Solid State Physics, ETH Z\"urich, 8093 Z\"urich, Switzerland}

\author{A. Zheludev}
\homepage{http://www.neutron.ethz.ch/}
\affiliation{Laboratory for Solid State Physics, ETH Z\"urich, 8093 Z\"urich, Switzerland}

\author{S. Mishra}
\affiliation{Laboratoire National des Champs Magn\'etiques Intenses, LNCMI-CNRS (UPR3228),
	EMFL, Universit\'e \\ Grenoble Alpes, UPS and INSA Toulouse, Bo\^{i}te Postale 166, 38042 Grenoble Cedex 9, France} 

\author{I. Sheikin}
\affiliation{Laboratoire National des Champs Magn\'etiques Intenses, LNCMI-CNRS (UPR3228),
	EMFL, Universit\'e \\ Grenoble Alpes, UPS and INSA Toulouse, Bo\^{i}te Postale 166, 38042 Grenoble Cedex 9, France} 

\author{S. Krämer}
\affiliation{Laboratoire National des Champs Magn\'etiques Intenses, LNCMI-CNRS (UPR3228),
	EMFL, Universit\'e \\ Grenoble Alpes, UPS and INSA Toulouse, Bo\^{i}te Postale 166, 38042 Grenoble Cedex 9, France} 

\author{M. Horvatić}
\affiliation{Laboratoire National des Champs Magn\'etiques Intenses, LNCMI-CNRS (UPR3228),
	EMFL, Universit\'e \\ Grenoble Alpes, UPS and INSA Toulouse, Bo\^{i}te Postale 166, 38042 Grenoble Cedex 9, France}

\author{A. Gazizulina}
\affiliation{Helmholtz-Zentrum Berlin für Materialien und Energie, Hahn-Meitner Platz 1, 14109 Berlin, Germany}

\author{O. Prokhnenko}
\affiliation{Helmholtz-Zentrum Berlin für Materialien und Energie, Hahn-Meitner Platz 1, 14109 Berlin, Germany}

\date{\today}

\begin{abstract}
Magnetization, magnetic torque, neutron diffraction and NMR experiments are used to map out the $H$$-$$T$ phase diagram of the prototypical quasi-two-dimensional ferro-antiferromagnet \PbV in magnetic fields up to 27~T. When the field is applied perpendicular to the axis of magnetic anisotropy, a new magnetic state emerges through a discontinuous transition and persists in a narrow field range just below saturation. The measured NMR spectra suggest a complex and possibly incommensurate magnetic order in that regime.

\end{abstract}

% PACS, the Physics and Astronomy Classification Scheme.
%\keywords{Suggested keywords}%Use showkeys class option if keyword display desired

\maketitle
%\tableofcontents

\section{Introduction} \label{sec:Intro}

In Heisenberg antiferromagnets (AFMs) the standard paradigm for polarization by an external magnetic field is that of single-magnon Bose-Einstein condensation (BEC). \cite{Batyev1984} It can be argued that this essentially semi-classical picture breaks down in quantum spin systems with sufficiently strong competition between AFM and ferromagnetic (FM) interactions.\cite{Ueda2013} Just below the saturation field, conventional AF long-range order is replaced by some entirely different state with non-trivial symmetries and possibly a phase separation. Depending on the details of the spin Hamiltonian, a host of novel {\em pre-saturation} quantum phases may occur, including chiral order, spin density waves (SDW) and a variety of multipolar (spin-nematic) phases.\cite{Shannon2006,Hikihara2008,Sudan2009,Sato2013,Starykh2014} Some of these novel phases have indeed been found in real ferro-antiferromagnetic materials.\cite{Svistov2011,Orlova2017,Willenberg2016,Pregelj2015,Bhartiya2019, Kohama2019}

Non-trivial pre-saturation phases are mainly expected to occur when, due to frustration, the system is close to a FM-AFM transition in zero field.\cite{Ueda2013,Starykh2014} The focus of the present study is the layered vanadyl phosphate \PbV. It was initially thought to be a good realization of the $J_1$$-$$J_2$ square lattice model with nearest neighbor (nn) FM and next nearest neighbor (nnn) AFM interactions.\cite{Kaul2004,Kaul2005,Skoulatos2007,Skoulatos2009,Forster2013,Nath2009} In a recent inelastic scattering study\cite{Bettler2019} we have shown that while it indeed is very two-dimensional and has both FM and AFM exchange constants,  the material is a poor approximation of a square lattice.  The degree of frustration is smaller than initially thought, placing it far from a FM-AFM instability. Moreover, it appears rather ``classical'' in that the ordered moment $m\sim 0.7~\mu_\mathrm{B}$ in the colinear AF ground state is rather large for a quasi-two-dimensional compound. It is therefore rather surprising that \PbV shows a highly non-trivial and probably incommensurate pre-saturation phase, the observation of which we report in the present paper, along with the mapping out of the entire $H$$-$$T$ phase diagram.

\begin{figure}
\centering
\includegraphics[width=1.0\columnwidth]{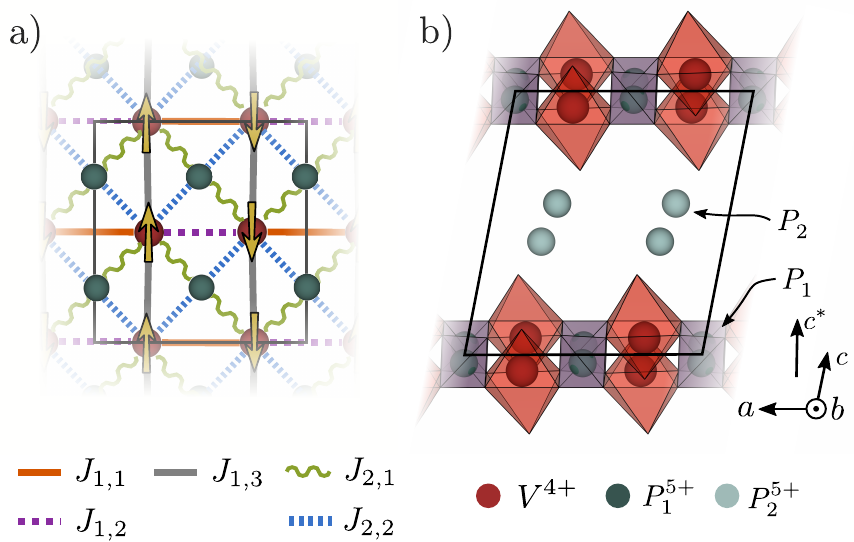}
\caption{Schematic view of the \PbV crystal structure. a) Single vanadophosphate layer in the $(a,b)$ crystallographic plane. Lines represent Heisenberg exchange constants between $S=1/2$ V$^{4+}$-ions, as described in the text. b) Projection of the entire structure onto the $(b,c)$ plane. There are two distinct inequivalent phosphorous sites $P_1$ and $P_2$. Shown are the VO$_5$ pyramids, the $P_1$O$_4$ tetrahedra and the $P_2$ sites. Oxygen and lead atoms are omitted for clarity.}
\label{fig::Crystal}
\end{figure}

The crystal structure of \PbV is shown in Fig.~\ref{fig::Crystal}. The unit cell  is monoclinic (P2$_{1}$/a)  with lattice parameters $a=8.747\text{~\AA}$, $b=9.016\text{~\AA}$, $c=9.863\text{~\AA}$  and $\beta=100.96^{\circ}$.\cite{Shpanchenko2006} The magnetism originates from the V$^{4+}$ ions, which host a $S=1/2$ spin and form two-dimensional layers in the $(a,b)$ plane. The four equivalent V$^{4+}$ sites in each unit cell are magnetically coupled by at least 5 inequivalent V-O-P-O-V superexchange pathways, as shown. The zero-field spin wave spectrum is adequately described by a Heisenberg hamiltonian with $J_{1,1}=-0.286(2)$~meV, $J_{1,2}=J_{1,3}=-0.389(2)$~meV, $J_{2,1}=+1.453(3)$~meV and $J_{2,2}=+0.538(2)$~meV.\cite{Bettler2019} In addition, there is a weak anisotropy with $\mathbf{b}$ as the magnetic easy axis. In zero field, long range order sets in at $T_N = 3.5$~K and corresponds to a collinear columnar-AF state (CAF), as indicated by arrows in Fig.~\ref{fig::Crystal}~(a). For a magnetic field along the crystallographic $\mathbf{b}$ direction, a conventional spin-flop transition occurs at $\mu_0H_\mathrm{SF}=0.94$~T.\cite{Kaul2005}  
%{\color{red} The magnetic saturation field is expected to exceed 20~T, which made it inaccessible in previous studies.} 
The magnetic saturation field is about 21~T,\cite{Tsirlin2009} which makes measurements experimentally challenging.
For the discussion below it is important to note the two inequivalent phosphorus sites, suitable for $^{31}$P NMR.\cite{Nath2009} The P$_{1}$ site lies in the V$^{4+}$ planes and is strongly coupled to their magnetism. In contrast, the coupling is much weaker for the interlayer P$_{2}$ site.

\begin{figure}[!t]
\centering
\includegraphics[width=1.0\linewidth]{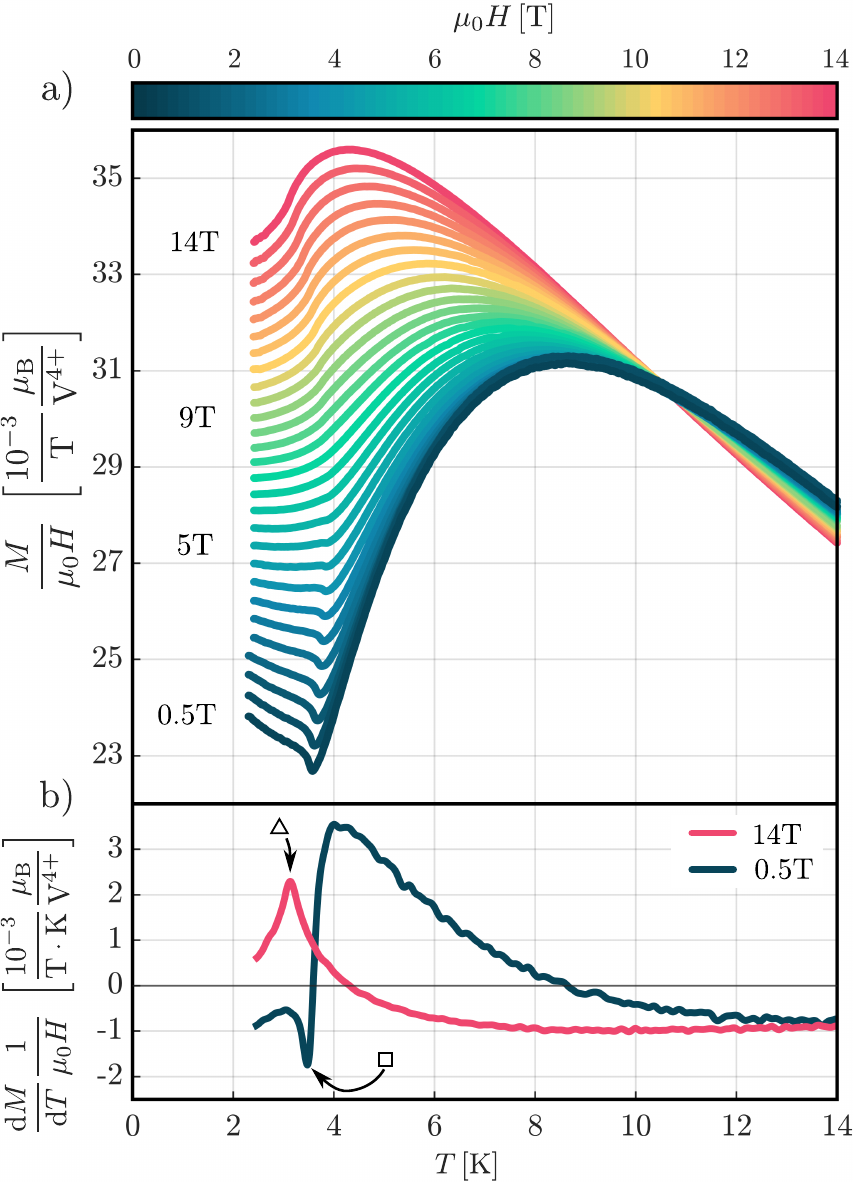}
\caption{a) Temperature dependence of magnetization measured in \PbV in different fields applied along the $\mathbf{c}^\ast$ axis. b) Typical derivatives of the curves shown in (a). Arrows indicate inflection points associated with the onset of magnetic long-range order. The symbols correspond to phase boundary points in Fig.~\ref{fig::diag}}.
\label{fig::mag}
\end{figure}

\section{Experimental} \label{sec:Crystal}

The main aim of this study was to investigate the $H-T$ phase diagram of \PbV in magnetic fields all the way to saturation. We used single crystal samples from the same batch as those used in Ref.~\onlinecite{Bettler2019}. Several complementary measurement techniques were employed. At low fields, the phase boundary was determined by magnetization measurements using a Quantum Design vibrating sample magnetometer (VSM) PPMS option. The external field was applied parallel to the $\mathbf{c}^\ast$ direction of a 43.6~mg sample. The data were collected while sweeping the temperature at constant fields. Fields up to 14~T and temperatures down to 2.5~K were explored.

%At the highest applied fields, the phase boundaries were traced in magnetic torque experiments 
Close to the saturation field, the phase boundaries were determined by means of capacitive torque magnetometry on a 0.4~mg sample. Two field geometries were explored, with the field applied very close to the $b$ and $c^*$ axes. The experiment was carried out at LNCMI-CNRS in Grenoble in a resistive magnet and with a dilution refrigerator equipped with an in-situ rotator. 
%{\color{green} Simon suggest to drop this part (I think he is right):} {\it{The sample was placed on a cantilever, which gets bent by the torque, caused by the external homogeneous magnetic field. The bending was detected by measuring the capacitance between the cantilever and the ground plate. The capacitance, which is assumed to be proportional to the torque, was measured by a mechanical capacitance bridge and a lock-in amplifier.}}

Some of the most insightful data were obtained in NMR experiments. These were carried out using a 21.4~mg sample in two different setups. For high temperatures ($T>1.4$~K) we employed a superconducting magnet with a regular $^4$He cryostat. Low temperature and high field data were collected using a resistive magnet and a dilution refrigerator. All measurements were carried out at LNCMI-CNRS in Grenoble. The field was in all cases applied along the crystalographic $\mathbf{c}^\ast$ direction and the NMR coil was parallel to $\mathbf{a}$. The spectra were obtained by the Fourier transforms of the $^{31}$P ($\gamma_N/2\pi = 17.236$~MHz/T) spin echo signals. Nuclear spin-lattice relaxation times ($T_1$) were measured by saturation-recovery experiments.
\begin{figure}[t]
\centering
\includegraphics[width=1.0\columnwidth]{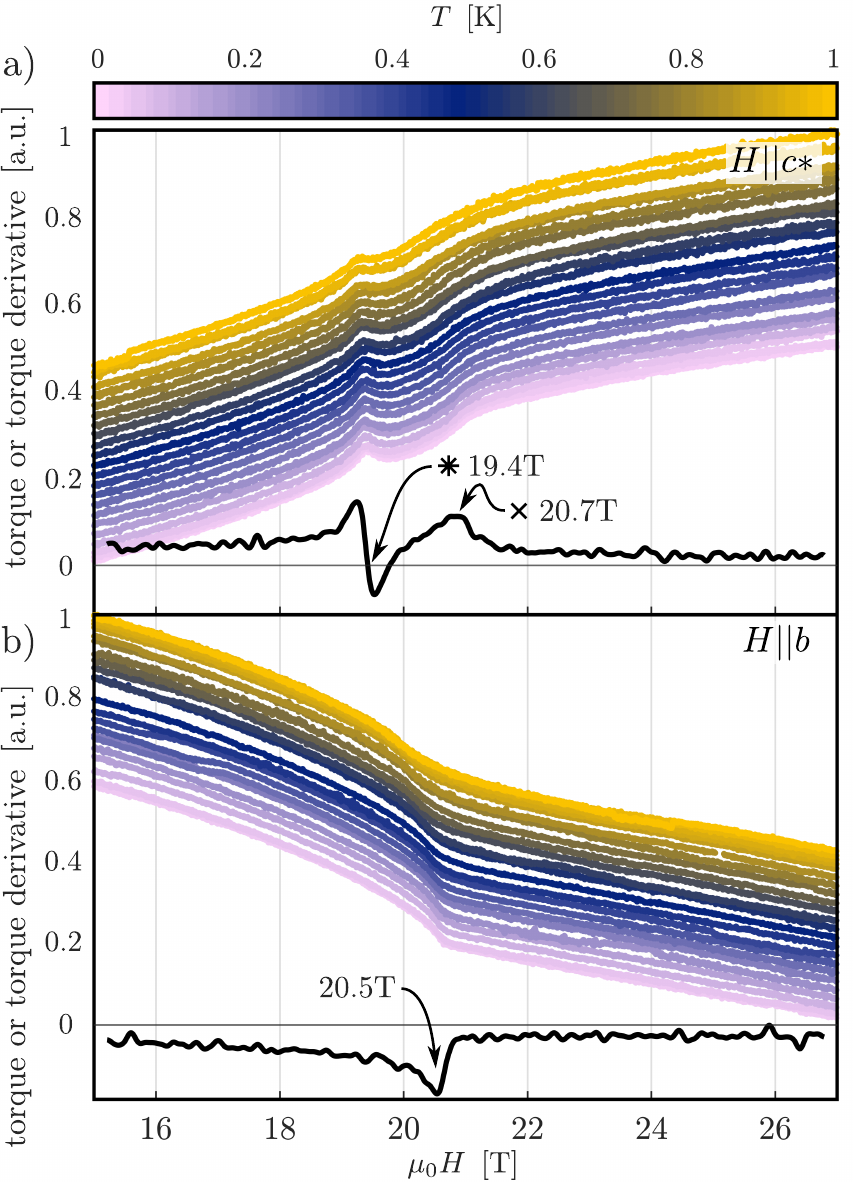}
\caption{Magnetic torque  measured in \PbV in magnetic fields applied along the $\mathbf{c}^\ast$ (a) and $\mathbf{b}$ (b) directions (color lines). The data collected at different temperatures are vertically offset for visibility. The black lines are field derivatives of the lowest temperature torque signal. The torque signal shows an inflection point at saturation (cross) and an extremum  at $H_c$ (asterisk). }
\label{fig::tor}
\end{figure}

\begin{figure}[!t]
\centering
\includegraphics[width=1.0\columnwidth]{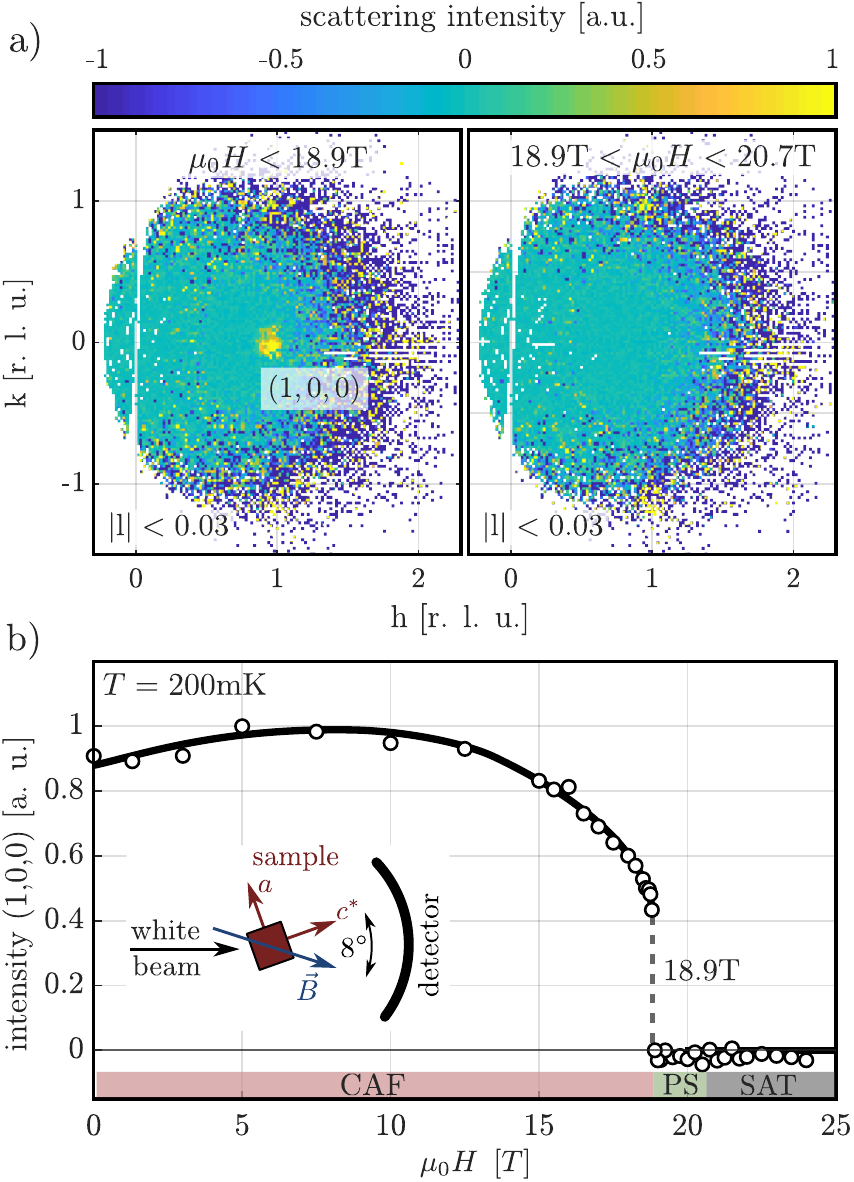}
\caption{a) A false color plot of the magnetic scattering intensity measured at $H<H_c$ (left) and for $H_c<H<H_\mathrm{sat}$ (right), as described in the text. b) Measured field dependence of the $(1,0,0)$ magnetic Bragg peak intensity in \PbV (circles). The inset illustrates the scattering geometry (angles not to scale). Lines are guides for the eye. }
\label{fig::diff}
\end{figure}

%Neutron diffraction experiments were repeated in ultra high magnetic fields  
Additional ultra high field neutron diffraction experiments were carried out with a 23~mg sample on the EXED time-of-flight instrument at HZB, in a  hybrid magnet providing dc fields up to 26~T.\cite{EXED2017} The experiment was performed at the base temperature of a dilution refrigerator ($T=200$~mK). The severely restrictive magnet geometry did not allow for an in-situ sample rotation.  In order to have the $(1,0,0)$ magnetic Bragg reflection in the narrow reciprocal space field of view of the instrument, we chose to apply the magnetic field at a fixed angle of 8$^{\circ}$ with respect to the $\mathbf{c}^\ast$ direction. The scattering geometry is shown in the inset of Fig.~\ref{fig::diff}~(a). The scattering intensity at each field value was measured with a counting time of roughly 60~min and was normalized to the spectral intensity of the incident white beam. To isolate out the magnetic contribution to scattering, the background was measured at $T=5$~K and zero magnetic field and subtracted from each data set.

\section{Experimental Results}

\subsubsection{Magnetization}
Magnetization measured in \PbV for $\mathbf{H}\|\mathbf{c}^\ast$ is plotted as a function of temperature in Fig.~\ref{fig::mag}~(a). Fig.~\ref{fig::mag}~(b) shows typical temperature derivatives computed for $\mu_0 H=0.5$ and $14$~T.  The phase transition is marked by inflection points in the $M(T)$ curves. At low field, the ordering temperature increases with field. In that regime the transition appears as a sharp dip in $\dd M/ \dd T$ [Fig.~\ref{fig::mag}~(b)]. Above about 7~T the trend is reversed, and the transition corresponds to a maximum of the derivative. The thus obtained phase boundary is plotted in open squares and triangles in  Fig.~\ref{fig::diag}. These results complement previous low-field studies in the $\mathbf{H}\|\mathbf{b}$ geometry.\cite{Kaul2005}

\subsubsection{Magnetic torque}
\begin{figure*}[t]
	\centering
	\includegraphics[width=1.0\linewidth]{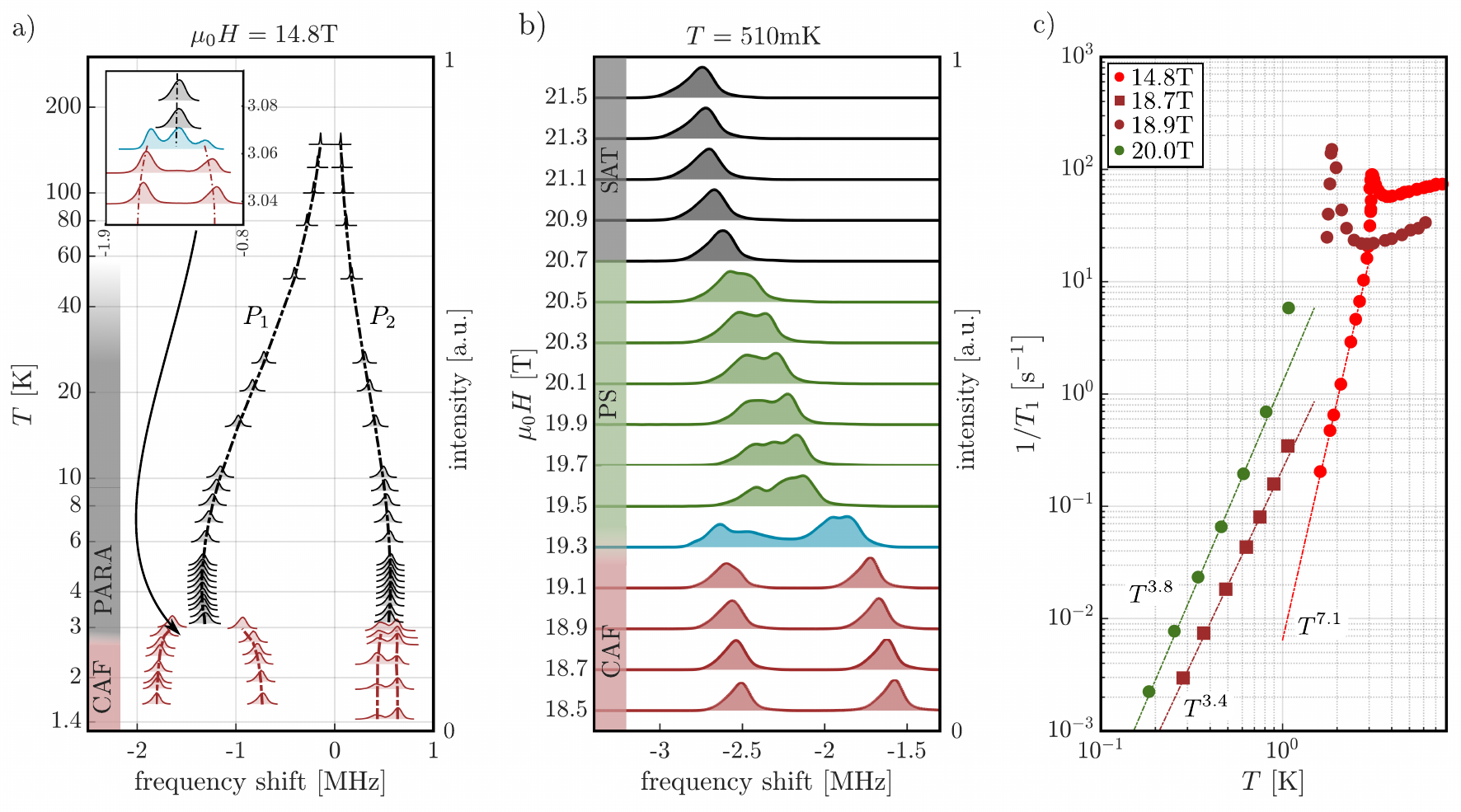}
	\caption{a) $^{31}$P-NMR spectra measured in \PbV at $\mu_0 H=14.8~T$ applied along $\mathbf{c}^\ast$. Data collected at different temperatures are offset according to the left (temperature) axis. Components ascribed to the two inequivalent phosphorous sites, P$_1$ and P$_2$ are labeled accordingly. b) $^{31}$P-NMR spectra of the P$_1$ site measured at $T=510$~mK.  Data collected  at different fields applied along $\mathbf{c}^\ast$ are offset according to the left (field) axis. Three magnetic phases as well as an apparent phase coexistence can be identified and the spectra are color-coded accordingly. c) Temperature dependence of the $^{31}$P spin-lattice relaxation rate $T_1^{-1}$  measured at several fields applied along $\mathbf{c}^\ast$ (symbols). Lines are power law fits to the data.}
	\label{fig::NMR}
\end{figure*}
The phase boundary at high field was deduced from magnetic torque data. The entire data set collected for two orientations is shown in Fig.~\ref{fig::tor}. Each line is vertically shifted for visibility and color-coded according to the temperature. 
%{\color{red}In both geometries the saturation field ($H_\mathrm{sat}$) is about 21~T.  It corresponds to an inflection point of the measured  torque curve (an extremum in the derivative), but appears more pronounced for $\mathbf{H}\| \mathbf{b}$ (cross). }
The saturation field correspond to an inflection point in the measured torque curve (to an extremum in the derivative) and is about 20.7~T and 20.5~T for $\mathbf{H}\|\mathbf{c^*}$ and $\mathbf{H}\|\mathbf{b}$ respectively. The peak in the field derivative, marked with a cross, appears more pronounced in the $\mathbf{H}\|\mathbf{b}$ geometry.
One of the main observations in this study is the additional sharp peak-like feature at $H_c\sim 19.4$~T for $\mathbf{H}\|\mathbf{c}^\ast$ (asterisk). We ascribe it to the emergence of an additional narrow pre-saturation phase (PS). For the $\mathbf{H}\|\mathbf{c}^\ast$-geometry, the temperature dependence of both features is plotted in the phase diagram~(Fig.~\ref{fig::diag}). A false-color plot of the torque signal derivative is shown in the background.

\subsubsection{Neutron diffraction}
The $(1,0,0)$ Bragg peak is the only magnetic reflection of the CAF phase accessible in the very restrictive high-field setup. The scattered magnetic intensity measured at $T=200$~mK integrated over the range $0.87<h<1.01$, $-0.08<k<0.08$ and $-0.01<l<0.025$ is plotted against field applied at 8$^\circ$ from $\mathbf{c}^\ast$ in Fig.~\ref{fig::diff}~(b).  The magnetic reflection disappears at $\mu_0H_c$~=~18.9~T in, what clearly appears to be, a discontinuous transition. Between $H_c$ and $H_\mathrm{sat}$ no other magnetic Bragg peaks could be located in the available data. This is borne out in Fig.~\ref{fig::diff}~(a), where we compare the average of all magnetic scattering  data collected for $H<H_c$ ($\mu_0 H=$0, 1.3, 3, 5, 7.5, 10, 12.5, 15, 15.5, 16, 16.5, 17, 17.5, 18, 18.25, 18.5, 18.6, 18.7, 18.75 and 18.8~T) and those for $H_c<H<H_\mathrm{sat}$ ( $\mu_0 H=$18.9, 19, 19.1, 19.25, 19.5, 19.75, 20, 20.25 and 20.5~T). Clearly no new strong peaks are present in the narrow slice  $|l|<0.03$ that these data sets correspond to. 
 Even tough the background data set, measured at 5~K in zero field, was subtracted from all data sets, the nuclear positions (2,0,0),(+1,1,0) and (-1,1,0) seem to show some scattering intensity. This is however an artefact of the bad counting statistic at large momentum transfer and the strong scattering at these positions. The incident beam has only a weak intensity for neutrons with high enough momentum to reach these positions resulting in large noise which gets further amplified by the normalization to the white beam. This noise is cropped of by the color scale which is chosen to make the magnetic (1,0,0) peak visible.

\subsubsection{NMR}
Fig. \ref{fig::NMR}~(a) shows the Fourier transform of the $^{31}$P NMR signal as a function of frequency shift (relative to the Larmor frequency, $\gamma_N H$) for various temperatures at a constant field of $\mu_0 H=14.8$~T. Each spectrum is normalized to its amplitude and vertically shifted according to its temperature. At high temperature two peaks are visible, corresponding to the two inequivalent phosphorous sites P$_1$ and P$_2$. The variation of their line-positions reflects the temperature dependence of the spin polarization. For the lower frequency line this variation is much stronger, so that this line is assigned to the P$_1$ site that is closer and thus more strongly coupled to V$^{4+}$ spins (Fig.~\ref{fig::Crystal}~(b) and Ref. \onlinecite{Nath2009}). Both lines split at $T$~= 3.06~K reflecting the staggered moment, that is the order parameter (OP), in the CAF phase. The insert in Fig.~\ref{fig::NMR}~(a) shows that both split and unsplit NMR lines coexistent at the transition temperature, meaning that a phase mixture exists in a very narrow temperature range. Furthermore, the line-splitting (OP) does not grow from zero but appears discontinuously at the transition, confirming that the transition is a discontinuous one. Nevertheless, in Fig.~\ref{fig::NMR}~(c) we see that the NMR $T_1^{-1}$ data, reflecting the low-energy spin dynamics/fluctuations, clearly display a sharp peak originating from the 3D critical fluctuations. This is characteristic of a {\it{second}}-order phase transition, as would be expected for a purely spin system. The transition is thus nearly second order (weakly first order), as predicted for a spin system that presents some magneto-elastic coupling. \cite{Amore2009,Wulf2015}

\begin{figure}
	\centering
	\includegraphics[width=1.0\linewidth]{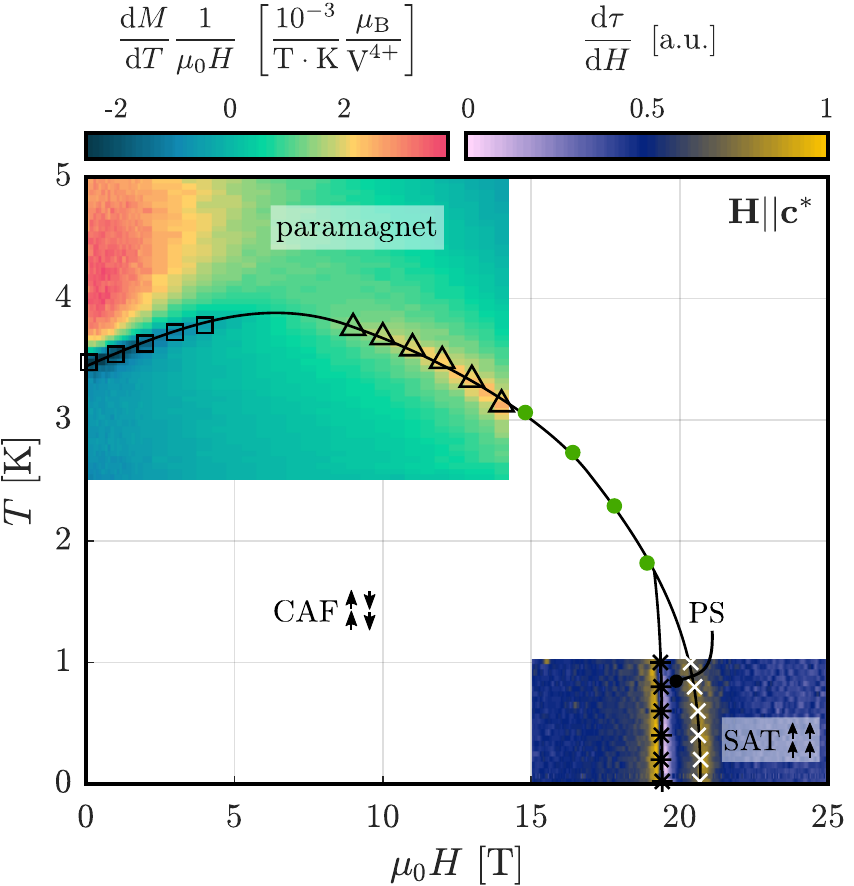}
	\caption{Combined magnetic phase diagram of \PbV in a magnetic field applied along the $\mathbf{c}^\ast$ direction. The false color plots are the derivative of magnetization with respect to temperature (top left) and the derivative of the torque signal with respect to field (bottom right). As described in the text, open squares and triangles are from magnetization data. Asterisks and crosses are from magnetic torque measurements. Solid green circles are from splitting of NMR lines, as shown in Fig. \ref{fig::NMR}~(a). Solid lines are guides for the eye. The columnar antiferromagnetic (CAF), paramagnetic and pre-saturation (PS) phases are labeled accordingly.}
	\label{fig::diag}
\end{figure}

To characterize the PS phase, Fig.~\ref{fig::NMR}~(b) shows the field dependence of the NMR spectra for the P$_1$ site taken at $T=510$~mK in the field range 18.5 - 21.5~T. Each spectrum is normalized to its amplitude and offset vertically according to the field value. The two NMR lines that characterize the OP of the CAF order merge at $H_c\sim 19.4$~T into a single broad structured peak. The latter persists, and gradually shrinks all the way to saturation at $H_\mathrm{sat} \sim 20.7$~T. The spectrum measured close to $H_c$ is indicative of a phase separation, in line with our expectations for a discontinuous transition.

Additional information on the ordered phases is provided by the temperature dependence of NMR $T_1^{-1}$ data, taken at $\mu_0 H=14.8$~T, 18.7/18.9~T and 20.0~T [Fig.~\ref{fig::NMR}~(c)]. The transition into the CAF phase presents a distinct cusp, reflecting critical spin fluctuations. In the ordered phases, the relaxation time follows a power law $T_1^{-1} \propto T^{\alpha}$ in a considerable temperature range. Deep in the CAF phase, the fitted exponent $\alpha$(14.8~T)~=~7.1(3) is very large, which is typical of high-order relaxation processes \cite{Beeman1968} observed in BEC-type systems \cite{Mayaffre2000, Jeong2017}. Close to $H_c$, the exponent is about two times smaller $\alpha$(18.7~T)~=~3.4(2), and very close to what is observed inside the PS phase $\alpha$(20.0~T)~=~3.8(2). This indicates that the spin dynamics of the CAF phase close to $H_c$ and of the PS phase are most probably similar, although the two phases present quite different static order.

\section{Discussion}
The measurements described above allow us to reconstruct the magnetic phase diagram for \PbV. For the case of a magnetic field applied along the easy axis of magnetic anisotropy ($\mathbf{H}\|\mathbf{b}$) it is as reported in Ref.~\onlinecite{Kaul2005}: A CAF phase with all spins along $\mathbf{b}$ in low field\cite{Bettler2019} followed by a CAF spin-flop phase all the way to the saturation. For this geometry, the only new result of the present study is a direct measurement of the saturation field $H_\mathrm{sat}\sim 20.5$~T, as deduced from our torque data.

Our main new finding is that, for a magnetic field applied perpendicular to the anisotropy axis, there is a new pre-saturation phase evident in the torque, neutron diffraction as well as NMR data. The corresponding phase diagram for $\mathbf{H}\|\mathbf{c}^\ast$ is shown in  Fig.~\ref{fig::diag}. From the neutron experiment, but also from the apparent presence of a mixed phase regime in NMR spectra, we conclude that the transition to the new phase is a discontinuous one. The critical field $H_c \approx 19.4$~T observed in the torque and NMR experiments are consistent. The discrepancy to $H_c\approx 18.9$~T obtained in the diffraction experiment is, most likely, due to the slightly different field orientation. As mentioned above, a discontinuous transition to non-trivial order can indeed be expected.\cite{Ueda2013} In the neutron data, we detect no sign of magnetic Bragg reflections between $H_c$ and $H_\mathrm{sat}$, but that may easily be due to the extremely restricted experimentally accessible region of reciprocal space. It is unlikely that the new phase is a spin-nematic, since NMR clearly shows a splitting or broadening of the resonance line. In one way or another, dipolar order must be  present. Furthermore, the structured shape of the NMR line hints at a complex and possibly incommensurate structure. As mentioned in the introduction, helimagnetic or SDW phases are the most likely candidates.

\section{Conclusion}
The appearance of a non-trivial incommensurate pre-saturation phase in a frustrated quantum ferro-antiferromagnet is in itself not surprising. However, that this phenomenon occurs in a system with a rather weak degree of frustration such as \PbV is quite remarkable.

\section{ACKNOWLEDGMENTS}

We acknowledge the support of the LNCMI-CNRS, member of the European Magnetic Field Laboratory (EMFL), and the Helmholtz Zentrum Berlin (HZB). 
The authors thank the magnet and sample environment team of HZB for the secure and smooth operation of the high field instrument EXED, these are S.Kempfer, P. Heller, P. Smeibidl and S. Gerischer R. Wahle respectively.
Work at ETH was supported by the Swiss National Science Foundation, Division 2.
\\

%\bibliography{../../../../../bib/azbib}
\bibliography{azbib}

\begin{thebibliography}{28}%
\makeatletter
\providecommand \@ifxundefined [1]{%
 \@ifx{#1\undefined}
}%
\providecommand \@ifnum [1]{%
 \ifnum #1\expandafter \@firstoftwo
 \else \expandafter \@secondoftwo
 \fi
}%
\providecommand \@ifx [1]{%
 \ifx #1\expandafter \@firstoftwo
 \else \expandafter \@secondoftwo
 \fi
}%
\providecommand \natexlab [1]{#1}%
\providecommand \enquote  [1]{``#1''}%
\providecommand \bibnamefont  [1]{#1}%
\providecommand \bibfnamefont [1]{#1}%
\providecommand \citenamefont [1]{#1}%
\providecommand \href@noop [0]{\@secondoftwo}%
\providecommand \href [0]{\begingroup \@sanitize@url \@href}%
\providecommand \@href[1]{\@@startlink{#1}\@@href}%
\providecommand \@@href[1]{\endgroup#1\@@endlink}%
\providecommand \@sanitize@url [0]{\catcode `\\12\catcode `\$12\catcode
  `\&12\catcode `\#12\catcode `\^12\catcode `\_12\catcode `\%12\relax}%
\providecommand \@@startlink[1]{}%
\providecommand \@@endlink[0]{}%
\providecommand \url  [0]{\begingroup\@sanitize@url \@url }%
\providecommand \@url [1]{\endgroup\@href {#1}{\urlprefix }}%
\providecommand \urlprefix  [0]{URL }%
\providecommand \Eprint [0]{\href }%
\providecommand \doibase [0]{http://dx.doi.org/}%
\providecommand \selectlanguage [0]{\@gobble}%
\providecommand \bibinfo  [0]{\@secondoftwo}%
\providecommand \bibfield  [0]{\@secondoftwo}%
\providecommand \translation [1]{[#1]}%
\providecommand \BibitemOpen [0]{}%
\providecommand \bibitemStop [0]{}%
\providecommand \bibitemNoStop [0]{.\EOS\space}%
\providecommand \EOS [0]{\spacefactor3000\relax}%
\providecommand \BibitemShut  [1]{\csname bibitem#1\endcsname}%
\let\auto@bib@innerbib\@empty
%</preamble>
\bibitem [{\citenamefont {Batyev}\ and\ \citenamefont
  {Braginski}(1984)}]{Batyev1984}%
  \BibitemOpen
  \bibfield  {author} {\bibinfo {author} {\bibfnamefont {E.~G.}\ \bibnamefont
  {Batyev}}\ and\ \bibinfo {author} {\bibfnamefont {L.~S.}\ \bibnamefont
  {Braginski}},\ }\href@noop {} {\bibfield  {journal} {\bibinfo  {journal}
  {Sov. Phys. JETP}\ }\textbf {\bibinfo {volume} {60}},\ \bibinfo {pages} {781}
  (\bibinfo {year} {1984})}\BibitemShut {NoStop}%
\bibitem [{\citenamefont {Ueda}\ and\ \citenamefont {Momoi}(2013)}]{Ueda2013}%
  \BibitemOpen
  \bibfield  {author} {\bibinfo {author} {\bibfnamefont {H.~T.}\ \bibnamefont
  {Ueda}}\ and\ \bibinfo {author} {\bibfnamefont {T.}~\bibnamefont {Momoi}},\
  }\href {\doibase 10.1103/PhysRevB.87.144417} {\bibfield  {journal} {\bibinfo
  {journal} {Phys. Rev. B}\ }\textbf {\bibinfo {volume} {87}},\ \bibinfo
  {pages} {144417} (\bibinfo {year} {2013})}\BibitemShut {NoStop}%
\bibitem [{\citenamefont {Shannon}\ \emph {et~al.}(2006)\citenamefont
  {Shannon}, \citenamefont {Momoi},\ and\ \citenamefont
  {Sindzingre}}]{Shannon2006}%
  \BibitemOpen
  \bibfield  {author} {\bibinfo {author} {\bibfnamefont {N.}~\bibnamefont
  {Shannon}}, \bibinfo {author} {\bibfnamefont {T.}~\bibnamefont {Momoi}}, \
  and\ \bibinfo {author} {\bibfnamefont {P.}~\bibnamefont {Sindzingre}},\
  }\href {\doibase 10.1103/PhysRevLett.96.027213} {\bibfield  {journal}
  {\bibinfo  {journal} {Phys. Rev. Lett.}\ }\textbf {\bibinfo {volume} {96}},\
  \bibinfo {pages} {027213} (\bibinfo {year} {2006})}\BibitemShut {NoStop}%
\bibitem [{\citenamefont {Hikihara}\ \emph {et~al.}(2008)\citenamefont
  {Hikihara}, \citenamefont {Kecke}, \citenamefont {Momoi},\ and\ \citenamefont
  {Furusaki}}]{Hikihara2008}%
  \BibitemOpen
  \bibfield  {author} {\bibinfo {author} {\bibfnamefont {T.}~\bibnamefont
  {Hikihara}}, \bibinfo {author} {\bibfnamefont {L.}~\bibnamefont {Kecke}},
  \bibinfo {author} {\bibfnamefont {T.}~\bibnamefont {Momoi}}, \ and\ \bibinfo
  {author} {\bibfnamefont {A.}~\bibnamefont {Furusaki}},\ }\href {\doibase
  10.1103/PhysRevB.78.144404} {\bibfield  {journal} {\bibinfo  {journal} {Phys.
  Rev. B}\ }\textbf {\bibinfo {volume} {78}},\ \bibinfo {pages} {144404}
  (\bibinfo {year} {2008})}\BibitemShut {NoStop}%
\bibitem [{\citenamefont {Sudan}\ \emph {et~al.}(2009)\citenamefont {Sudan},
  \citenamefont {L\"uscher},\ and\ \citenamefont {L\"auchli}}]{Sudan2009}%
  \BibitemOpen
  \bibfield  {author} {\bibinfo {author} {\bibfnamefont {J.}~\bibnamefont
  {Sudan}}, \bibinfo {author} {\bibfnamefont {A.}~\bibnamefont {L\"uscher}}, \
  and\ \bibinfo {author} {\bibfnamefont {A.~M.}\ \bibnamefont {L\"auchli}},\
  }\href {\doibase 10.1103/PhysRevB.80.140402} {\bibfield  {journal} {\bibinfo
  {journal} {Phys. Rev. B}\ }\textbf {\bibinfo {volume} {80}},\ \bibinfo
  {pages} {140402(R)} (\bibinfo {year} {2009})}\BibitemShut {NoStop}%
\bibitem [{\citenamefont {Sato}\ \emph {et~al.}(2013)\citenamefont {Sato},
  \citenamefont {Hikihara},\ and\ \citenamefont {Momoi}}]{Sato2013}%
  \BibitemOpen
  \bibfield  {author} {\bibinfo {author} {\bibfnamefont {M.}~\bibnamefont
  {Sato}}, \bibinfo {author} {\bibfnamefont {T.}~\bibnamefont {Hikihara}}, \
  and\ \bibinfo {author} {\bibfnamefont {T.}~\bibnamefont {Momoi}},\ }\href
  {\doibase 10.1103/PhysRevLett.110.077206} {\bibfield  {journal} {\bibinfo
  {journal} {Phys. Rev. Lett.}\ }\textbf {\bibinfo {volume} {110}},\ \bibinfo
  {pages} {077206} (\bibinfo {year} {2013})}\BibitemShut {NoStop}%
\bibitem [{\citenamefont {Starykh}\ and\ \citenamefont
  {Balents}(2014)}]{Starykh2014}%
  \BibitemOpen
  \bibfield  {author} {\bibinfo {author} {\bibfnamefont {O.~A.}\ \bibnamefont
  {Starykh}}\ and\ \bibinfo {author} {\bibfnamefont {L.}~\bibnamefont
  {Balents}},\ }\href {\doibase 10.1103/PhysRevB.89.104407} {\bibfield
  {journal} {\bibinfo  {journal} {Phys. Rev. B}\ }\textbf {\bibinfo {volume}
  {89}},\ \bibinfo {pages} {104407} (\bibinfo {year} {2014})}\BibitemShut
  {NoStop}%
\bibitem [{\citenamefont {Svistov}\ \emph {et~al.}(2011)\citenamefont
  {Svistov}, \citenamefont {Fujita}, \citenamefont {Yamaguchi}, \citenamefont
  {Kimura}, \citenamefont {Omura}, \citenamefont {Prokofiev}, \citenamefont
  {Smirnov}, \citenamefont {Honda},\ and\ \citenamefont
  {Hagiwara}}]{Svistov2011}%
  \BibitemOpen
  \bibfield  {author} {\bibinfo {author} {\bibfnamefont {L.~E.}\ \bibnamefont
  {Svistov}}, \bibinfo {author} {\bibfnamefont {T.}~\bibnamefont {Fujita}},
  \bibinfo {author} {\bibfnamefont {H.}~\bibnamefont {Yamaguchi}}, \bibinfo
  {author} {\bibfnamefont {S.}~\bibnamefont {Kimura}}, \bibinfo {author}
  {\bibfnamefont {K.}~\bibnamefont {Omura}}, \bibinfo {author} {\bibfnamefont
  {A.}~\bibnamefont {Prokofiev}}, \bibinfo {author} {\bibfnamefont {A.~I.}\
  \bibnamefont {Smirnov}}, \bibinfo {author} {\bibfnamefont {Z.}~\bibnamefont
  {Honda}}, \ and\ \bibinfo {author} {\bibfnamefont {M.}~\bibnamefont
  {Hagiwara}},\ }\href {\doibase 10.1134/S0021364011010073} {\bibfield
  {journal} {\bibinfo  {journal} {JETP Lett.}\ }\textbf {\bibinfo {volume}
  {93}},\ \bibinfo {pages} {21} (\bibinfo {year} {2011})}\BibitemShut {NoStop}%
\bibitem [{\citenamefont {Orlova}\ \emph {et~al.}(2017)\citenamefont {Orlova},
  \citenamefont {Green}, \citenamefont {Law}, \citenamefont {Gorbunov},
  \citenamefont {Chanda}, \citenamefont {Kr\"amer}, \citenamefont
  {Horvati\ifmmode~\acute{c}\else \'{c}\fi{}}, \citenamefont {Kremer},
  \citenamefont {Wosnitza},\ and\ \citenamefont {Rikken}}]{Orlova2017}%
  \BibitemOpen
  \bibfield  {author} {\bibinfo {author} {\bibfnamefont {A.}~\bibnamefont
  {Orlova}}, \bibinfo {author} {\bibfnamefont {E.~L.}\ \bibnamefont {Green}},
  \bibinfo {author} {\bibfnamefont {J.~M.}\ \bibnamefont {Law}}, \bibinfo
  {author} {\bibfnamefont {D.~I.}\ \bibnamefont {Gorbunov}}, \bibinfo {author}
  {\bibfnamefont {G.}~\bibnamefont {Chanda}}, \bibinfo {author} {\bibfnamefont
  {S.}~\bibnamefont {Kr\"amer}}, \bibinfo {author} {\bibfnamefont
  {M.}~\bibnamefont {Horvati\ifmmode~\acute{c}\else \'{c}\fi{}}}, \bibinfo
  {author} {\bibfnamefont {R.~K.}\ \bibnamefont {Kremer}}, \bibinfo {author}
  {\bibfnamefont {J.}~\bibnamefont {Wosnitza}}, \ and\ \bibinfo {author}
  {\bibfnamefont {G.~L. J.~A.}\ \bibnamefont {Rikken}},\ }\href {\doibase
  10.1103/PhysRevLett.118.247201} {\bibfield  {journal} {\bibinfo  {journal}
  {Phys. Rev. Lett.}\ }\textbf {\bibinfo {volume} {118}},\ \bibinfo {pages}
  {247201} (\bibinfo {year} {2017})}\BibitemShut {NoStop}%
\bibitem [{\citenamefont {Willenberg}\ \emph {et~al.}(2016)\citenamefont
  {Willenberg}, \citenamefont {Sch\"apers}, \citenamefont {Wolter},
  \citenamefont {Drechsler}, \citenamefont {Reehuis}, \citenamefont {Hoffmann},
  \citenamefont {B\"uchner}, \citenamefont {Studer}, \citenamefont {Rule},
  \citenamefont {Ouladdiaf}, \citenamefont {S\"ullow},\ and\ \citenamefont
  {Nishimoto}}]{Willenberg2016}%
  \BibitemOpen
  \bibfield  {author} {\bibinfo {author} {\bibfnamefont {B.}~\bibnamefont
  {Willenberg}}, \bibinfo {author} {\bibfnamefont {M.}~\bibnamefont
  {Sch\"apers}}, \bibinfo {author} {\bibfnamefont {A.~U.~B.}\ \bibnamefont
  {Wolter}}, \bibinfo {author} {\bibfnamefont {S.-L.}\ \bibnamefont
  {Drechsler}}, \bibinfo {author} {\bibfnamefont {M.}~\bibnamefont {Reehuis}},
  \bibinfo {author} {\bibfnamefont {J.-U.}\ \bibnamefont {Hoffmann}}, \bibinfo
  {author} {\bibfnamefont {B.}~\bibnamefont {B\"uchner}}, \bibinfo {author}
  {\bibfnamefont {A.~J.}\ \bibnamefont {Studer}}, \bibinfo {author}
  {\bibfnamefont {K.~C.}\ \bibnamefont {Rule}}, \bibinfo {author}
  {\bibfnamefont {B.}~\bibnamefont {Ouladdiaf}}, \bibinfo {author}
  {\bibfnamefont {S.}~\bibnamefont {S\"ullow}}, \ and\ \bibinfo {author}
  {\bibfnamefont {S.}~\bibnamefont {Nishimoto}},\ }\href {\doibase
  10.1103/PhysRevLett.116.047202} {\bibfield  {journal} {\bibinfo  {journal}
  {Phys. Rev. Lett.}\ }\textbf {\bibinfo {volume} {116}},\ \bibinfo {pages}
  {047202} (\bibinfo {year} {2016})}\BibitemShut {NoStop}%
\bibitem [{\citenamefont {Pregelj}\ \emph {et~al.}(2015)\citenamefont
  {Pregelj}, \citenamefont {Zorko}, \citenamefont {Zaharko}, \citenamefont
  {Nojiri}, \citenamefont {Berger}, \citenamefont {Chapon},\ and\ \citenamefont
  {Arčon}}]{Pregelj2015}%
  \BibitemOpen
  \bibfield  {author} {\bibinfo {author} {\bibfnamefont {M.}~\bibnamefont
  {Pregelj}}, \bibinfo {author} {\bibfnamefont {A.}~\bibnamefont {Zorko}},
  \bibinfo {author} {\bibfnamefont {O.}~\bibnamefont {Zaharko}}, \bibinfo
  {author} {\bibfnamefont {H.}~\bibnamefont {Nojiri}}, \bibinfo {author}
  {\bibfnamefont {H.}~\bibnamefont {Berger}}, \bibinfo {author} {\bibfnamefont
  {L.~C.}\ \bibnamefont {Chapon}}, \ and\ \bibinfo {author} {\bibfnamefont
  {D.}~\bibnamefont {Arčon}},\ }\href {\doibase 10.1038/ncomms8255} {\bibfield
   {journal} {\bibinfo  {journal} {Nat. Comm.}\ }\textbf {\bibinfo {volume}
  {6}},\ \bibinfo {pages} {7255} (\bibinfo {year} {2015})}\BibitemShut
  {NoStop}%
\bibitem [{\citenamefont {Bhartiya}\ \emph {et~al.}(2019)\citenamefont
  {Bhartiya}, \citenamefont {Povarov}, \citenamefont {Blosser}, \citenamefont
  {Bettler}, \citenamefont {Yan}, \citenamefont {Gvasaliya}, \citenamefont
  {Raymond}, \citenamefont {Ressouche}, \citenamefont {Beauvois}, \citenamefont
  {Xu}, \citenamefont {Yokaichiya},\ and\ \citenamefont
  {Zheludev}}]{Bhartiya2019}%
  \BibitemOpen
  \bibfield  {author} {\bibinfo {author} {\bibfnamefont {V.~K.}\ \bibnamefont
  {Bhartiya}}, \bibinfo {author} {\bibfnamefont {K.~Y.}\ \bibnamefont
  {Povarov}}, \bibinfo {author} {\bibfnamefont {D.}~\bibnamefont {Blosser}},
  \bibinfo {author} {\bibfnamefont {S.}~\bibnamefont {Bettler}}, \bibinfo
  {author} {\bibfnamefont {Z.}~\bibnamefont {Yan}}, \bibinfo {author}
  {\bibfnamefont {S.}~\bibnamefont {Gvasaliya}}, \bibinfo {author}
  {\bibfnamefont {S.}~\bibnamefont {Raymond}}, \bibinfo {author} {\bibfnamefont
  {E.}~\bibnamefont {Ressouche}}, \bibinfo {author} {\bibfnamefont
  {K.}~\bibnamefont {Beauvois}}, \bibinfo {author} {\bibfnamefont
  {J.}~\bibnamefont {Xu}}, \bibinfo {author} {\bibfnamefont {F.}~\bibnamefont
  {Yokaichiya}}, \ and\ \bibinfo {author} {\bibfnamefont {A.}~\bibnamefont
  {Zheludev}},\ }\href {\doibase 10.1103/PhysRevResearch.1.033078} {\bibfield
  {journal} {\bibinfo  {journal} {Phys. Rev. Research}\ }\textbf {\bibinfo
  {volume} {1}},\ \bibinfo {pages} {033078} (\bibinfo {year}
  {2019})}\BibitemShut {NoStop}%
\bibitem [{\citenamefont {Kohama}\ \emph {et~al.}(2019)\citenamefont {Kohama},
  \citenamefont {Ishikawa}, \citenamefont {Matsuo}, \citenamefont {Kindo},
  \citenamefont {Shannon},\ and\ \citenamefont {Hiroi}}]{Kohama2019}%
  \BibitemOpen
  \bibfield  {author} {\bibinfo {author} {\bibfnamefont {Y.}~\bibnamefont
  {Kohama}}, \bibinfo {author} {\bibfnamefont {H.}~\bibnamefont {Ishikawa}},
  \bibinfo {author} {\bibfnamefont {A.}~\bibnamefont {Matsuo}}, \bibinfo
  {author} {\bibfnamefont {K.}~\bibnamefont {Kindo}}, \bibinfo {author}
  {\bibfnamefont {N.}~\bibnamefont {Shannon}}, \ and\ \bibinfo {author}
  {\bibfnamefont {Z.}~\bibnamefont {Hiroi}},\ }\href {\doibase
  10.1073/pnas.1821969116} {\bibfield  {journal} {\bibinfo  {journal}
  {Proceedings of the National Academy of Sciences}\ }\textbf {\bibinfo
  {volume} {116}},\ \bibinfo {pages} {10686} (\bibinfo {year} {2019})},\
  \Eprint
  {http://arxiv.org/abs/https://www.pnas.org/content/116/22/10686.full.pdf}
  {https://www.pnas.org/content/116/22/10686.full.pdf} \BibitemShut {NoStop}%
\bibitem [{\citenamefont {Kaul}\ \emph {et~al.}(2004)\citenamefont {Kaul},
  \citenamefont {Rosner}, \citenamefont {Shannon}, \citenamefont
  {Shpanchenko},\ and\ \citenamefont {Geibel}}]{Kaul2004}%
  \BibitemOpen
  \bibfield  {author} {\bibinfo {author} {\bibfnamefont {E.}~\bibnamefont
  {Kaul}}, \bibinfo {author} {\bibfnamefont {H.}~\bibnamefont {Rosner}},
  \bibinfo {author} {\bibfnamefont {N.}~\bibnamefont {Shannon}}, \bibinfo
  {author} {\bibfnamefont {R.}~\bibnamefont {Shpanchenko}}, \ and\ \bibinfo
  {author} {\bibfnamefont {C.}~\bibnamefont {Geibel}},\ }\href {\doibase
  https://doi.org/10.1016/j.jmmm.2003.12.002} {\bibfield  {journal} {\bibinfo
  {journal} {Journal of Magnetism and Magnetic Materials}\ }\textbf {\bibinfo
  {volume} {272-276}},\ \bibinfo {pages} {922 } (\bibinfo {year} {2004})},\
  \bibinfo {note} {proceedings of the International Conference on Magnetism
  (ICM 2003)}\BibitemShut {NoStop}%
\bibitem [{\citenamefont {Kaul}(2005)}]{Kaul2005}%
  \BibitemOpen
  \bibfield  {author} {\bibinfo {author} {\bibfnamefont {E.~E.}\ \bibnamefont
  {Kaul}},\ }\emph {\bibinfo {title} {Experimental Investigation of New
  Low-Dimensional Spin Systems in Vanadium Oxides}},\ \href@noop {} {Ph.D.
  thesis},\ \bibinfo  {school} {Technische Universit\"at Dresden} (\bibinfo
  {year} {2005})\BibitemShut {NoStop}%
\bibitem [{\citenamefont {Skoulatos}\ \emph {et~al.}(2007)\citenamefont
  {Skoulatos}, \citenamefont {Goff}, \citenamefont {Shannon}, \citenamefont
  {Kaul}, \citenamefont {Geibel}, \citenamefont {Murani}, \citenamefont
  {Enderle},\ and\ \citenamefont {Wildes}}]{Skoulatos2007}%
  \BibitemOpen
  \bibfield  {author} {\bibinfo {author} {\bibfnamefont {M.}~\bibnamefont
  {Skoulatos}}, \bibinfo {author} {\bibfnamefont {J.}~\bibnamefont {Goff}},
  \bibinfo {author} {\bibfnamefont {N.}~\bibnamefont {Shannon}}, \bibinfo
  {author} {\bibfnamefont {E.}~\bibnamefont {Kaul}}, \bibinfo {author}
  {\bibfnamefont {C.}~\bibnamefont {Geibel}}, \bibinfo {author} {\bibfnamefont
  {A.}~\bibnamefont {Murani}}, \bibinfo {author} {\bibfnamefont
  {M.}~\bibnamefont {Enderle}}, \ and\ \bibinfo {author} {\bibfnamefont
  {A.}~\bibnamefont {Wildes}},\ }\href {\doibase
  https://doi.org/10.1016/j.jmmm.2006.10.379} {\bibfield  {journal} {\bibinfo
  {journal} {Journal of Magnetism and Magnetic Materials}\ }\textbf {\bibinfo
  {volume} {310}},\ \bibinfo {pages} {1257 } (\bibinfo {year} {2007})},\
  \bibinfo {note} {proceedings of the 17th International Conference on
  Magnetism}\BibitemShut {NoStop}%
\bibitem [{\citenamefont {Skoulatos}\ \emph {et~al.}(2009)\citenamefont
  {Skoulatos}, \citenamefont {Goff}, \citenamefont {Geibel}, \citenamefont
  {Kaul}, \citenamefont {Nath}, \citenamefont {Shannon}, \citenamefont
  {Schmidt}, \citenamefont {Murani}, \citenamefont {Deen}, \citenamefont
  {Enderle},\ and\ \citenamefont {Wildes}}]{Skoulatos2009}%
  \BibitemOpen
  \bibfield  {author} {\bibinfo {author} {\bibfnamefont {M.}~\bibnamefont
  {Skoulatos}}, \bibinfo {author} {\bibfnamefont {J.~P.}\ \bibnamefont {Goff}},
  \bibinfo {author} {\bibfnamefont {C.}~\bibnamefont {Geibel}}, \bibinfo
  {author} {\bibfnamefont {E.~E.}\ \bibnamefont {Kaul}}, \bibinfo {author}
  {\bibfnamefont {R.}~\bibnamefont {Nath}}, \bibinfo {author} {\bibfnamefont
  {N.}~\bibnamefont {Shannon}}, \bibinfo {author} {\bibfnamefont
  {B.}~\bibnamefont {Schmidt}}, \bibinfo {author} {\bibfnamefont {A.~P.}\
  \bibnamefont {Murani}}, \bibinfo {author} {\bibfnamefont {P.~P.}\
  \bibnamefont {Deen}}, \bibinfo {author} {\bibfnamefont {M.}~\bibnamefont
  {Enderle}}, \ and\ \bibinfo {author} {\bibfnamefont {A.~R.}\ \bibnamefont
  {Wildes}},\ }\href {http://stacks.iop.org/0295-5075/88/i=5/a=57005}
  {\bibfield  {journal} {\bibinfo  {journal} {EPL (Europhysics Letters)}\
  }\textbf {\bibinfo {volume} {88}},\ \bibinfo {pages} {57005} (\bibinfo {year}
  {2009})}\BibitemShut {NoStop}%
\bibitem [{\citenamefont {F\"orster}\ \emph {et~al.}(2013)\citenamefont
  {F\"orster}, \citenamefont {Garcia}, \citenamefont {Gruner}, \citenamefont
  {Kaul}, \citenamefont {Schmidt}, \citenamefont {Geibel},\ and\ \citenamefont
  {Sichelschmidt}}]{Forster2013}%
  \BibitemOpen
  \bibfield  {author} {\bibinfo {author} {\bibfnamefont {T.}~\bibnamefont
  {F\"orster}}, \bibinfo {author} {\bibfnamefont {F.~A.}\ \bibnamefont
  {Garcia}}, \bibinfo {author} {\bibfnamefont {T.}~\bibnamefont {Gruner}},
  \bibinfo {author} {\bibfnamefont {E.~E.}\ \bibnamefont {Kaul}}, \bibinfo
  {author} {\bibfnamefont {B.}~\bibnamefont {Schmidt}}, \bibinfo {author}
  {\bibfnamefont {C.}~\bibnamefont {Geibel}}, \ and\ \bibinfo {author}
  {\bibfnamefont {J.}~\bibnamefont {Sichelschmidt}},\ }\href {\doibase
  10.1103/PhysRevB.87.180401} {\bibfield  {journal} {\bibinfo  {journal} {Phys.
  Rev. B}\ }\textbf {\bibinfo {volume} {87}},\ \bibinfo {pages} {180401}
  (\bibinfo {year} {2013})}\BibitemShut {NoStop}%
\bibitem [{\citenamefont {Nath}\ \emph {et~al.}(2009)\citenamefont {Nath},
  \citenamefont {Furukawa}, \citenamefont {Borsa}, \citenamefont {Kaul},
  \citenamefont {Baenitz}, \citenamefont {Geibel},\ and\ \citenamefont
  {Johnston}}]{Nath2009}%
  \BibitemOpen
  \bibfield  {author} {\bibinfo {author} {\bibfnamefont {R.}~\bibnamefont
  {Nath}}, \bibinfo {author} {\bibfnamefont {Y.}~\bibnamefont {Furukawa}},
  \bibinfo {author} {\bibfnamefont {F.}~\bibnamefont {Borsa}}, \bibinfo
  {author} {\bibfnamefont {E.~E.}\ \bibnamefont {Kaul}}, \bibinfo {author}
  {\bibfnamefont {M.}~\bibnamefont {Baenitz}}, \bibinfo {author} {\bibfnamefont
  {C.}~\bibnamefont {Geibel}}, \ and\ \bibinfo {author} {\bibfnamefont {D.~C.}\
  \bibnamefont {Johnston}},\ }\href {\doibase 10.1103/PhysRevB.80.214430}
  {\bibfield  {journal} {\bibinfo  {journal} {Phys. Rev. B}\ }\textbf {\bibinfo
  {volume} {80}},\ \bibinfo {pages} {214430} (\bibinfo {year}
  {2009})}\BibitemShut {NoStop}%
\bibitem [{\citenamefont {Bettler}\ \emph {et~al.}(2019)\citenamefont
  {Bettler}, \citenamefont {Landolt}, \citenamefont {Aksoy}, \citenamefont
  {Yan}, \citenamefont {Gvasaliya}, \citenamefont {Qiu}, \citenamefont
  {Ressouche}, \citenamefont {Beauvois}, \citenamefont {Raymond}, \citenamefont
  {Ponomaryov}, \citenamefont {Zvyagin},\ and\ \citenamefont
  {Zheludev}}]{Bettler2019}%
  \BibitemOpen
  \bibfield  {author} {\bibinfo {author} {\bibfnamefont {S.}~\bibnamefont
  {Bettler}}, \bibinfo {author} {\bibfnamefont {F.}~\bibnamefont {Landolt}},
  \bibinfo {author} {\bibfnamefont {O.~M.}\ \bibnamefont {Aksoy}}, \bibinfo
  {author} {\bibfnamefont {Z.}~\bibnamefont {Yan}}, \bibinfo {author}
  {\bibfnamefont {S.}~\bibnamefont {Gvasaliya}}, \bibinfo {author}
  {\bibfnamefont {Y.}~\bibnamefont {Qiu}}, \bibinfo {author} {\bibfnamefont
  {E.}~\bibnamefont {Ressouche}}, \bibinfo {author} {\bibfnamefont
  {K.}~\bibnamefont {Beauvois}}, \bibinfo {author} {\bibfnamefont
  {S.}~\bibnamefont {Raymond}}, \bibinfo {author} {\bibfnamefont {A.~N.}\
  \bibnamefont {Ponomaryov}}, \bibinfo {author} {\bibfnamefont {S.~A.}\
  \bibnamefont {Zvyagin}}, \ and\ \bibinfo {author} {\bibfnamefont
  {A.}~\bibnamefont {Zheludev}},\ }\href {\doibase 10.1103/PhysRevB.99.184437}
  {\bibfield  {journal} {\bibinfo  {journal} {Phys. Rev. B}\ }\textbf {\bibinfo
  {volume} {99}},\ \bibinfo {pages} {184437} (\bibinfo {year}
  {2019})}\BibitemShut {NoStop}%
\bibitem [{\citenamefont {Shpanchenko}\ \emph {et~al.}(2006)\citenamefont
  {Shpanchenko}, \citenamefont {Kaul}, \citenamefont {Geibel},\ and\
  \citenamefont {Antipov}}]{Shpanchenko2006}%
  \BibitemOpen
  \bibfield  {author} {\bibinfo {author} {\bibfnamefont {V.}~\bibnamefont
  {Shpanchenko}}, \bibinfo {author} {\bibfnamefont {E.~E.}\ \bibnamefont
  {Kaul}}, \bibinfo {author} {\bibfnamefont {C.}~\bibnamefont {Geibel}}, \ and\
  \bibinfo {author} {\bibfnamefont {E.~V.}\ \bibnamefont {Antipov}},\
  }\href@noop {} {\bibfield  {journal} {\bibinfo  {journal} {Acta Crystallog.
  C}\ }\textbf {\bibinfo {volume} {62}},\ \bibinfo {pages} {88} (\bibinfo
  {year} {2006})}\BibitemShut {NoStop}%
\bibitem [{\citenamefont {Tsirlin}\ \emph {et~al.}(2009)\citenamefont
  {Tsirlin}, \citenamefont {Schmidt}, \citenamefont {Skourski}, \citenamefont
  {Nath}, \citenamefont {Geibel},\ and\ \citenamefont {Rosner}}]{Tsirlin2009}%
  \BibitemOpen
  \bibfield  {author} {\bibinfo {author} {\bibfnamefont {A.~A.}\ \bibnamefont
  {Tsirlin}}, \bibinfo {author} {\bibfnamefont {B.}~\bibnamefont {Schmidt}},
  \bibinfo {author} {\bibfnamefont {Y.}~\bibnamefont {Skourski}}, \bibinfo
  {author} {\bibfnamefont {R.}~\bibnamefont {Nath}}, \bibinfo {author}
  {\bibfnamefont {C.}~\bibnamefont {Geibel}}, \ and\ \bibinfo {author}
  {\bibfnamefont {H.}~\bibnamefont {Rosner}},\ }\href {\doibase
  10.1103/PhysRevB.80.132407} {\bibfield  {journal} {\bibinfo  {journal} {Phys.
  Rev. B}\ }\textbf {\bibinfo {volume} {80}},\ \bibinfo {pages} {132407}
  (\bibinfo {year} {2009})}\BibitemShut {NoStop}%
\bibitem [{\citenamefont {Prokhnenko}\ \emph {et~al.}(2017)\citenamefont
  {Prokhnenko}, \citenamefont {Smeibidl}, \citenamefont {Stein}, \citenamefont
  {Bartkowiak},\ and\ \citenamefont {St\"usser}}]{EXED2017}%
  \BibitemOpen
  \bibfield  {author} {\bibinfo {author} {\bibfnamefont {O.}~\bibnamefont
  {Prokhnenko}}, \bibinfo {author} {\bibfnamefont {P.}~\bibnamefont
  {Smeibidl}}, \bibinfo {author} {\bibfnamefont {W.-D.}\ \bibnamefont {Stein}},
  \bibinfo {author} {\bibfnamefont {M.}~\bibnamefont {Bartkowiak}}, \ and\
  \bibinfo {author} {\bibfnamefont {N.}~\bibnamefont {St\"usser}},\ }\href
  {\doibase 10.17815/jlsrf-3-111} {\bibfield  {journal} {\bibinfo  {journal}
  {Journal of large-scale research facilities {JLSRF}}\ }\textbf {\bibinfo
  {volume} {3}} (\bibinfo {year} {2017}),\ 10.17815/jlsrf-3-111}\BibitemShut
  {NoStop}%
\bibitem [{\citenamefont {Dell'Amore}\ \emph {et~al.}(2009)\citenamefont
  {Dell'Amore}, \citenamefont {Schilling},\ and\ \citenamefont
  {Kr\"amer}}]{Amore2009}%
  \BibitemOpen
  \bibfield  {author} {\bibinfo {author} {\bibfnamefont {R.}~\bibnamefont
  {Dell'Amore}}, \bibinfo {author} {\bibfnamefont {A.}~\bibnamefont
  {Schilling}}, \ and\ \bibinfo {author} {\bibfnamefont {K.}~\bibnamefont
  {Kr\"amer}},\ }\href {\doibase 10.1103/PhysRevB.79.014438} {\bibfield
  {journal} {\bibinfo  {journal} {Phys. Rev. B}\ }\textbf {\bibinfo {volume}
  {79}},\ \bibinfo {pages} {014438} (\bibinfo {year} {2009})}\BibitemShut
  {NoStop}%
\bibitem [{\citenamefont {Wulf}\ \emph {et~al.}(2015)\citenamefont {Wulf},
  \citenamefont {H\"uvonen}, \citenamefont {Sch\"onemann}, \citenamefont
  {K\"uhne}, \citenamefont {Herrmannsd\"orfer}, \citenamefont {Glavatskyy},
  \citenamefont {Gerischer}, \citenamefont {Kiefer}, \citenamefont
  {Gvasaliya},\ and\ \citenamefont {Zheludev}}]{Wulf2015}%
  \BibitemOpen
  \bibfield  {author} {\bibinfo {author} {\bibfnamefont {E.}~\bibnamefont
  {Wulf}}, \bibinfo {author} {\bibfnamefont {D.}~\bibnamefont {H\"uvonen}},
  \bibinfo {author} {\bibfnamefont {R.}~\bibnamefont {Sch\"onemann}}, \bibinfo
  {author} {\bibfnamefont {H.}~\bibnamefont {K\"uhne}}, \bibinfo {author}
  {\bibfnamefont {T.}~\bibnamefont {Herrmannsd\"orfer}}, \bibinfo {author}
  {\bibfnamefont {I.}~\bibnamefont {Glavatskyy}}, \bibinfo {author}
  {\bibfnamefont {S.}~\bibnamefont {Gerischer}}, \bibinfo {author}
  {\bibfnamefont {K.}~\bibnamefont {Kiefer}}, \bibinfo {author} {\bibfnamefont
  {S.}~\bibnamefont {Gvasaliya}}, \ and\ \bibinfo {author} {\bibfnamefont
  {A.}~\bibnamefont {Zheludev}},\ }\href {\doibase 10.1103/PhysRevB.91.014406}
  {\bibfield  {journal} {\bibinfo  {journal} {Phys. Rev. B}\ }\textbf {\bibinfo
  {volume} {91}},\ \bibinfo {pages} {014406} (\bibinfo {year}
  {2015})}\BibitemShut {NoStop}%
\bibitem [{\citenamefont {Beeman}\ and\ \citenamefont
  {Pincus}(1968)}]{Beeman1968}%
  \BibitemOpen
  \bibfield  {author} {\bibinfo {author} {\bibfnamefont {D.}~\bibnamefont
  {Beeman}}\ and\ \bibinfo {author} {\bibfnamefont {P.}~\bibnamefont
  {Pincus}},\ }\href {\doibase 10.1103/PhysRev.166.359} {\bibfield  {journal}
  {\bibinfo  {journal} {Phys. Rev.}\ }\textbf {\bibinfo {volume} {166}},\
  \bibinfo {pages} {359} (\bibinfo {year} {1968})}\BibitemShut {NoStop}%
\bibitem [{\citenamefont {Mayaffre}\ \emph {et~al.}(2000)\citenamefont
  {Mayaffre}, \citenamefont {Horvati\ifmmode~\acute{c}\else \'{c}\fi{}},
  \citenamefont {Berthier}, \citenamefont {Julien}, \citenamefont
  {S\'egransan}, \citenamefont {L\'evy},\ and\ \citenamefont
  {Piovesana}}]{Mayaffre2000}%
  \BibitemOpen
  \bibfield  {author} {\bibinfo {author} {\bibfnamefont {H.}~\bibnamefont
  {Mayaffre}}, \bibinfo {author} {\bibfnamefont {M.}~\bibnamefont
  {Horvati\ifmmode~\acute{c}\else \'{c}\fi{}}}, \bibinfo {author}
  {\bibfnamefont {C.}~\bibnamefont {Berthier}}, \bibinfo {author}
  {\bibfnamefont {M.-H.}\ \bibnamefont {Julien}}, \bibinfo {author}
  {\bibfnamefont {P.}~\bibnamefont {S\'egransan}}, \bibinfo {author}
  {\bibfnamefont {L.}~\bibnamefont {L\'evy}}, \ and\ \bibinfo {author}
  {\bibfnamefont {O.}~\bibnamefont {Piovesana}},\ }\href {\doibase
  10.1103/PhysRevLett.85.4795} {\bibfield  {journal} {\bibinfo  {journal}
  {Phys. Rev. Lett.}\ }\textbf {\bibinfo {volume} {85}},\ \bibinfo {pages}
  {4795} (\bibinfo {year} {2000})}\BibitemShut {NoStop}%
\bibitem [{\citenamefont {Jeong}\ \emph {et~al.}(2017)\citenamefont {Jeong},
  \citenamefont {Mayaffre}, \citenamefont {Berthier}, \citenamefont
  {Schmidiger}, \citenamefont {Zheludev},\ and\ \citenamefont
  {Horvati\ifmmode~\acute{c}\else \'{c}\fi{}}}]{Jeong2017}%
  \BibitemOpen
  \bibfield  {author} {\bibinfo {author} {\bibfnamefont {M.}~\bibnamefont
  {Jeong}}, \bibinfo {author} {\bibfnamefont {H.}~\bibnamefont {Mayaffre}},
  \bibinfo {author} {\bibfnamefont {C.}~\bibnamefont {Berthier}}, \bibinfo
  {author} {\bibfnamefont {D.}~\bibnamefont {Schmidiger}}, \bibinfo {author}
  {\bibfnamefont {A.}~\bibnamefont {Zheludev}}, \ and\ \bibinfo {author}
  {\bibfnamefont {M.}~\bibnamefont {Horvati\ifmmode~\acute{c}\else
  \'{c}\fi{}}},\ }\href {\doibase 10.1103/PhysRevLett.118.167206} {\bibfield
  {journal} {\bibinfo  {journal} {Phys. Rev. Lett.}\ }\textbf {\bibinfo
  {volume} {118}},\ \bibinfo {pages} {167206} (\bibinfo {year}
  {2017})}\BibitemShut {NoStop}%
\end{thebibliography}%

\end{document}